\documentclass[12pt]{article}
\usepackage{graphicx}
\usepackage{amsfonts}
\usepackage{amssymb,amsmath}
\usepackage{latexsym}
\usepackage{color}
\usepackage{cite}
\input{colordvi.tex}

\renewcommand{\thefootnote}{\#\arabic{footnote}}

\begin{document}
\setcounter{footnote}{0}

\begin{titlepage}

\begin{center}


\vskip .5in

{\Large \bf
Scale-dependent bias from the primordial non-Gaussianity with
a Gaussian-squared field }

\vskip .45in

{\large
Shuichiro Yokoyama
}

\vskip .1in

{shu@a.phys.nagoya-u.ac.jp}

\vskip .40in

{\em
Department of Physics and Astrophysics, Nagoya University,
Aichi 464-8602, Japan}

\end{center}

\vskip .4in

\begin{abstract}
We investigate the halo bias in the case
where the primordial curvature fluctuations, $\Phi$, 
are sourced from both a Gaussian random field and a Gaussian-squared field, as
$\Phi({\bf x}) = \phi({\bf x}) + \psi({\bf x})^2 - \langle \psi({\bf x})^2\rangle$,
so-called "ungaussiton model".
We employ the peak-background split formula
and 
find a new scale-dependence in the halo bias induced
from the Gaussian-squared field.
\end{abstract}
\end{titlepage}

\renewcommand{\thepage}{\arabic{page}}
\setcounter{page}{1}
\renewcommand{\thefootnote}{\#\arabic{footnote}}

\section{Introduction}
Primordial non-Gaussianity has been attracting attention as a new probe of the
physics of the early Universe, e.g., inflation models.
There are a lot of theoretical models of predicting the primordial curvature fluctuations
with the large non-Gaussian features and several types of the primordial non-Gaussianity
have been predicted (For recent reviews, see e.g. Ref.~\cite{Sasaki:2010sw}).

On the observational side,
precise measurements
of the cosmic microwave background (CMB) anisotropies are ones of the most powerful tools to
hunt for the primordial non-Gaussianity (see e.g. Refs.~\cite{Komatsu:2010hc,Komatsu:2010fb}).
Current CMB data indicates that the primordial adiabatic fluctuations
follow nearly perfect Gaussianity. However, there still remains the possibility of
detecting the non-Gaussianity in the future experiments such as Planck~\cite{:2006uk}.
In the CMB experiments,
the non-Gaussianity could be detected by non-zero higher order correlation functions such as
the three-point function (bispectrum) and the four-point one (trispectrum).

Recently, the observations of the large-scale structure (LSS) of the Universe
have been also focused on as other powerful tools to detect the primordial non-Gaussianity
just as the CMB observations (see e.g. Refs.~\cite{Verde:2010wp,Desjacques:2010nn}).
In particular, it is known that the primordial non-Gaussianity induces the
modification of the halo mass function for more massive objects at higher
redshift~(see e.g. Refs.~\cite{Matarrese:2000iz,LoVerde:2007ri,DeSimone:2010mu})
and also a scale-dependence
of the halo bias~(see e.g. Refs. \cite{Dalal:2007cu,Matarrese:2008nc,Slosar:2008hx}). 
Since future surveys of the LSS will provide
a large amount of samples of galaxies over a huge volume,
it is expected that the LSS observations
would give a tighter constraint on the primordial non-Gaussianity
comparable to that obtained from the future CMB observations~\cite{Slosar:2008hx}.

In accordance with the recent progress of observations,
we are invited to consider the effect of the higher order non-Gaussianities, e.g.,
the non-zero four point correlation function of the primordial fluctuations
\cite{Desjacques:2009jb,Maggiore:2009hp,Chongchitnan:2010xz,Chongchitnan:2010hb,
Enqvist:2010bg,LoVerde:2011iz,Yokoyama:2011sy,Desjacques:2011jb,Desjacques:2011mq,Smith:2011ub,Gong:2011gx},
or that of the primordial non-Gaussianity in a multi-field inflationary model
where the primordial fluctuations are sourced from multi-field fluctuations~\cite{Tseliakhovich:2010kf,Smith:2010gx}.
Following these works, in this paper, 
we investigate the halo bias in
the case where the primordial adiabatic curvature fluctuations 
are given by~\cite{Boubekeur:2005fj,Suyama:2008nt}
\begin{eqnarray}
\Phi ({\bf x}) = 
\phi_{\rm G} ({\bf x}) + \psi_{\rm G} ({\bf x})^2 - \langle \psi_{\rm G} ({\bf x})^2\rangle~,
\label{eq:form}
\end{eqnarray}
where $\langle \phi \psi \rangle = 0$.
In Ref.~\cite{Suyama:2008nt}, such a model has been dubbed
"ungaussiton" model.
This type of the non-Gaussianity can be realized in the case where
the primordial fluctuations are sourced from both the inflaton and the curvaton
fluctuations and the curvaton stays at the origin during inflation.
In this case, 
due to the absence of the linear term of $\psi$
the non-zero bispectrum and trispectrum of the curvature perturbations
respectively come from the six- and eight-orders in $\psi$.
Hence,
this model predicts
a specific consistency relation between the bispectrum and the trispectrum
which would be confirmed in the CMB experiments
and it should be interesting to examine the effect of such a primordial non-Gaussianity
on the LSS surveys, in particular, the scale-dependence of the halo bias.
In order to obtain an analytic expression for the halo bias,
we employ the peak-background split formula~\cite{Dalal:2007cu,Slosar:2008hx}
which is a useful tool to calculate the halo bias
with the local-type non-Gaussianity.

This paper is organized as follows.
In the next section, we briefly review the so-called "ungaussiton" model
and show the expression for the bispectrum and the trispectrum of
the primordial curvature fluctuations in this model and also the consistency relation
between the bispectrum and the trispectrum.
In section~\ref{sec:bias}, we give an expression for the halo bias in this model
by making use of the peak-background formula.
Section~\ref{sec:sum} is devoted
to the summary and discussion.

\section{Non-Gaussianity in the primordial bi- and tri-spectrum in the "ungaussiton" model}
\label{sec:formula}

Here, we briefly review the so-called "ungaussiton" model~\cite{Boubekeur:2005fj,Suyama:2008nt}
and show some interesting
consequence of this model by considering the bispectrum and the trispectrum of the
primordial curvature perturbations.
This model can be realized in the case where
the primordial fluctuations are sourced from both the inflaton and the curvaton
fluctuations and the curvaton stays at the origin during inflation.
In this model where the primordial curvature fluctuations are given by Eq.~(\ref{eq:form}),
the power spectrum, the bispectrum and the trispectrum of the primordial curvature fluctuations
are respectively given by~\cite{Boubekeur:2005fj,Suyama:2008nt}
\begin{eqnarray}
\langle \Phi({\bf k}) \Phi ({\bf k}')\rangle &=& (2 \pi)^3 P_{\Phi}(k) 
\delta^{(3)} ({\bf k} + {\bf k}')~,\nonumber\\
P_\Phi(k) &=&
P_\phi(k) \left[
1 + 4 R^2 {\cal P}_\phi \ln \left( k_{\rm max} L \right) \right] ~, \label{eq:power} 
\end{eqnarray}
\begin{eqnarray}
\langle \Phi({\bf k}_1)
\Phi({\bf k}_2)
\Phi({\bf k}_3) 
\rangle &=&
(2 \pi )^3 B_\Phi(k_1,k_2,k_3) \delta^{(3)} ({\bf k}_1 + {\bf k}_2 + {\bf k}_3)~, \nonumber\\
B_\Phi(k_1,k_2,k_3) &=&
8R^3{\cal P}_\phi \ln \left( k_b L \right) \bigl[ P_\phi(k_1)P_\phi(k_2) \nonumber\\
&&
\qquad\qquad
+ P_\phi(k_2)P_\phi(k_3)
+ P_\phi(k_3)P_\phi(k_1)\bigr]~,\label{eq:bis}
\end{eqnarray}
and
\begin{eqnarray}
\langle \Phi({\bf k}_1)\Phi({\bf k}_2)\Phi({\bf k}_3)\Phi({\bf k}_4)\rangle_{\rm c}
&=&
(2\pi)^3 T_\Phi(k_1,k_2,k_3,k_4) \delta^{(3)} ({\bf k}_1 + {\bf k}_2 + {\bf k}_3 + {\bf k}_4)~, \nonumber\\
T_\Phi(k_1,k_2,k_3,k_4) &=&
16 R^4 {\cal P}_\phi \ln \left( k_t L \right) \nonumber\\
&&\qquad
\times
\bigl[ P_\phi(k_1)P_\phi(k_2)P_\phi(k_{13}) + 11~{\rm perms.} \bigr]~,\label{eq:tri}
\end{eqnarray}
where $L$ is the size of the box in which the fluctuations are observed,
${\cal P}_\phi = k^3 P_\phi(k) / (2 \pi^2)$, $R = P_\psi(k) / P_\phi(k)$, $k_b = {\rm min}\{ k_i \}$,
$k_t = {\rm min}\{ k_{ij}, k_\ell \}$, $k_{ij} = \left| {\bf k}_i + {\bf k}_j \right|$
and a subscript, ${\rm c}$, denotes the connected part.

Since the current observations indicate that the primordial curvature fluctuations are almost Gaussian,
the power spectrum of $\Phi$ should not be dominated by the second term
in the bracket of the right hand side in Eq.~(\ref{eq:power})
as
\begin{eqnarray}
4 R^2 {\cal P}_\phi \ln \left( k_{\rm max} L \right) < 1~.
\label{eq:condition}
\end{eqnarray}
By making use of 
the non-linearity parameters defined as~\cite{Boubekeur:2005fj,Komatsu:2001rj,Byrnes:2006vq}
\begin{eqnarray}
B_\Phi(k_1,k_2,k_3) &=&
2 f_{\rm NL} \bigl[ P_\phi(k_1)P_\phi(k_2) + + P_\phi(k_2)P_\phi(k_3)
+ P_\phi(k_3)P_\phi(k_1)\bigr] \cr
T_\Phi(k_1,k_2,k_3,k_4)
& = &
{25 \over 9} \tau_{\rm NL}
\bigl[ P_\phi(k_1)P_\phi(k_2)P_\phi(k_{13}) + 11~{\rm perms.} \bigr]~,
\end{eqnarray}
we have
\begin{eqnarray}
f_{\rm NL} &=& 4 R^3 {\cal P}_\phi \ln (k_b L)~, \cr\cr
{25 \over 9}\tau_{\rm NL} &=& 16 R^4 {\cal P}_\phi  \ln (k_t L)~.
\label{eq:ft}
\end{eqnarray}
From these expressions,
we find that the large non-linearity parameters can be realized even under the condition
given by Eq.~(\ref{eq:condition}) and
we can obtain a relation between the non-linearity parameters in this model, which is given by
\begin{eqnarray}
\tau_{\rm NL}
&=& \alpha \left( {6 \over 5} f_{\rm NL}\right)^{4/3} \left({3 \over 5}\right)^{2/3}
{\cal P}_{\phi}^{-1/3} \nonumber\\
&\sim&
0.8 \times 10^3 \times \alpha \left( {6 \over 5}f_{\rm NL} \right)^{4/3}~,
\label{eq:consistency}
\end{eqnarray}
where $\alpha \equiv \ln \left( k_t L\right) / \left( \ln \left(k_b L\right)\right)^{4/3}$
is a coefficient of order unity and we have used ${\cal P}_\phi \sim {\cal P}_\Phi \simeq 7.2 \times 10^{-10}$.
By using this relations,
we can distinguish this model from the others
which predict the large non-Gaussianity and the other relations
between the non-linearity parameters $f_{\rm NL}$ and $\tau_{\rm NL}$,
in particular, the models which give the standard consistency relation given by
$\tau_{\rm NL} = (6f_{\rm NL} / 5)^2$ for local-type non-Gaussianity. 

\section{Halo bias in peak-background split formalism}
\label{sec:bias}

Here, 
following Refs.~\cite{Dalal:2007cu,Slosar:2008hx},
we calculate the halo bias for the "ungaussiton" model where the primordial curvature perturbations
are given by Eq.~(\ref{eq:form}) in the context of the peak-background split formalism.

In the non-Gaussian case, the large and small scale density fluctuations are not independent.
Let us decompose the primordial curvature perturbations into the long- and short-wavelength parts as
\begin{eqnarray}
\Phi ({\bf x}) &=& \phi_l({\bf x}) + \phi_s ({\bf x})
+
\left[
\left( \psi_l({\bf x}) + \psi_s({\bf x})
\right)^2 - \langle \psi_l({\bf x})^2 \rangle - \langle \psi_s({\bf x})^2 \rangle \right] \nonumber\\
&=&
\phi_l({\bf x}) +
\psi_l({\bf x})^2 
- \langle \psi_l({\bf x})^2 \rangle \nonumber\\
 &&
+ \phi_s ({\bf x})
+
2 \psi_l({\bf x})\psi_s({\bf x})+ \psi_s({\bf x})^2
 - \langle \psi_s({\bf x})^2 \rangle
~,
\label{eq:split}
\end{eqnarray}
where we have assumed $\langle \phi({\bf x}) \psi({\bf x}) \rangle = 0$ and
the long- and short-wavelength parts are uncorrelated.
From this equation, we can obtain expressions for the long- and short-wavelength modes of the density fields
in Fourier space as
\begin{eqnarray}
\delta_l({\bf k}) = {\cal M}(k) 
\left[
\phi_l({\bf k}) 
+
\int {d^3{\rm p} \over (2\pi)^3}
\psi_l({\bf p})\psi_l({\bf k} - {\bf p}) - 
(2\pi)^3 \delta^{(3)}({\bf k}) \langle \psi_l({\bf x})^2\rangle \right]
~,
\end{eqnarray}
and
\begin{eqnarray}
\delta_s({\bf k}, {\bf x}) = {\cal M}(k) \left[ \phi_s ({\bf k}) + 2 \psi_l({\bf x}) \psi_s({\bf k})\right]~,
\end{eqnarray}
with
\begin{eqnarray}
{\cal M}(k) = {2T(k) \over 3\Omega_{m0} H_0^2}~,
\end{eqnarray}
where we have neglected the contribution from the $\psi_s({\bf x})^2$ term
because it is known that such a quadratic term does not affect the halo bias~\cite{Slosar:2008hx}. 
Here, $T(k)$ is the matter transfer function, $\Omega_{m0}$ is the present matter density parameter
and $H_0$ is the present Hubble parameter.
From this equation, we can find that the non-Gaussianity affects the rescaling of the
amplitude of the density fluctuations on small scales as
\begin{eqnarray}
P_{\delta_s}({\bf x}) &=& {\cal M}(k)^2 \left[P_{\phi_s}
+ 4\psi_l({\bf x})^2 P_{\psi_s} \right] \cr\cr
&=& {\cal M}(k)^2 \left[ 1 + 4R \psi_l({\bf x})^2 \right]P_{\phi_s}~,
\end{eqnarray}
and hence
a standard cosmological parameter, $\sigma_8$, which denotes the rms of
the linear density field with $8h^{-1} {\rm Mpc}$ smoothing,
depends on the position ${\bf x}$ as
\begin{eqnarray}
\sigma_8 ({\bf x}) \approx \left[ 1 + 2 R \left(\psi_l({\bf x})^2 -  \langle\psi_l({\bf x})^2\rangle
\right) \right] \sigma_8~.
\end{eqnarray}
where we have introduced the $\langle \psi_l({\bf x})^2\rangle$ term
in order to achieve $\langle \sigma_8 ({\bf x}) \rangle = \sigma_8$.
%
Due to the long-wavelength modes of the density fluctuations and
also the above effect of the primordial non-Gaussianity,
the density of halos $n({\bf x})$ in a large box at position ${\bf x}$
deviates from the mean density $\bar{n}$.
Following Refs.~\cite{Slosar:2008hx,Tseliakhovich:2010kf,Smith:2010gx},
$n ({\bf x})$ is given by
\begin{eqnarray}
n ({\bf x}) = \bar{n} \left( 1 + \delta_l({\bf x}) \right)
\left(
1 + \delta_l({\bf x}) {\partial \log \bar{n} \over \partial \delta_l}
+ 2R \left(\psi_l({\bf x})^2 - \langle \psi_l({\bf x})^2\rangle\right) {\partial \log \bar{n} \over \partial \log \sigma_8}
\right)~,
\end{eqnarray}
where the $(1+\delta_l)$ comes from transforming Lagrangian to Eulerian space.
From this equation, we can obtain the density fluctuations of halos,
$\delta_h({\bf x}) \equiv (n ({\bf x}) - \bar{n})/\bar{n}$, as
\begin{eqnarray}
\delta_h({\bf x}) \approx \left( 1 +  {\partial \log \bar{n} \over \partial \delta_l}
\right)\delta_l({\bf x})
+ 2 R {\partial \log \bar{n} \over \partial \log \sigma_8}
\left(
\psi_l({\bf x})^2 - \langle \psi_l({\bf x})^2 \rangle
\right)~,
\end{eqnarray}
where we have drop the second order terms of $\delta_l$.
In the Fourier space, we have
\begin{eqnarray}
\delta_h({\bf k}) = b_0 \delta ({\bf k})
+ 2 R \delta_c \left( b_0 - 1 \right) \left[
\int {d^3{\bf p} \over (2\pi)^3}
\psi({\bf p})\psi({\bf k} - {\bf p}) - 
(2\pi)^3 \delta^{(3)}({\bf k}) \langle \psi({\bf x})^2\rangle
\right]~,
\label{eq:halodens}
\end{eqnarray}
where~\cite{Slosar:2008hx,Smith:2010gx}
\begin{eqnarray}
b_0 \equiv 1 + {\partial \log \bar{n} \over \partial \delta_l}~,
\end{eqnarray}
and we have used 
\begin{eqnarray}
{\partial \log \bar{n} \over \partial \log \sigma_8}
 = \delta_c {\partial \log \bar{n} \over \partial \delta_l}~,
\end{eqnarray}
with $\delta_c$ being a critical density.
The equation~(\ref{eq:halodens}) is one of the main result of this paper.
However, defining the halo bias from this equation is somewhat ambiguous.
Hence, let us consider the cross power spectrum of the density fluctuations of halos
and matter fluctuations
and also
the power spectrum of the density fluctuations of halos.
The cross power spectrum is given by
\begin{eqnarray}
\langle \delta_h({\bf k}) \delta ({\bf k}') \rangle &=& (2\pi)^3 P_{h\delta}(k)\delta^{(3)}({\bf k} + {\bf k}')~, \cr\cr
P_{h\delta}(k) &=& \left[
b_0 + 8 \delta_c (b_0 - 1) R^3 {\cal P}_\phi \ln (k_{\rm max} L ) \right] P_\delta(k) \cr\cr
&=&
\left[
b_0 + {2 \beta f_{\rm NL} \delta_c (b_0 - 1) \over {\cal M}(k)} \right] P_\delta(k)
~,
\end{eqnarray}
where we have used Eq.~(\ref{eq:ft}) and $\beta \equiv \ln (k_{\rm max}L) / \ln (k_b L)$.
Once the halo bias is defined as $b_h(k) \equiv P_{h\delta}(k)/P_\delta(k) $,
we can obtain the halo bias in the "ungaussiton" model as
\begin{eqnarray}
b_h(k,z) = b_0 + {2 \beta f_{\rm NL}\delta_c (b_0-1) \over {\cal M}(k)D(z)}~,
\end{eqnarray}
where we have introduced the linear growth function, $D(z)$.
Since a parameter $\beta$ is order of unity,
this expression is just corresponding to the one in the standard local-type primordial non-Gaussianity case
\cite{Slosar:2008hx}.

Let us consider the power spectrum of the density fluctuations of halos and it is given by
\begin{eqnarray}
\langle \delta_h({\bf k})\delta_h({\bf k}') \rangle 
&=&
(2\pi)^3P_h(k)\delta^{(3)}({\bf k} + {\bf k}') ~,\cr\cr
P_h(k) &=& \left[
b_0^2 + 2b_0{2 \beta f_{\rm NL} \delta_c (b_0 - 1) \over {\cal M}(k)}
+{ \gamma ( 25\tau_{\rm NL}/9 ) \delta_c^2 (b_0 - 1)^2 \over {\cal M}(k)^{2}}
\right] P_\delta(k)~, 
\end{eqnarray}
where we have used Eq.~(\ref{eq:ft}) and $\gamma \equiv \ln (k_{\rm max}L) / \ln (k_t L)$.
If the consistency relation between $f_{\rm NL}$ and $\tau_{\rm NL}$
is given by $\tau_{\rm NL} = 36 f_{\rm NL}^2 / 25$
with the assumption that $\beta = \gamma = 1$, then
$P_h(k)/P_\delta(k) = b_h(k)^2$ is realized.
It is well known that
this result should be realized in the standard local-type primordial non-Gaussianity case.
However, since in the "ungaussiton" model
the relation between $f_{\rm NL}$ and $\tau_{\rm NL}$ is
given by Eq.~(\ref{eq:consistency}),
$P_h(k) / P_\delta(k) = b_h(k)^2$ can not be realized any more.
Instead, we find that in the "ungaussiton" model
we have
\begin{eqnarray}
P_h(k)/P_\delta(k) =
\left[
b_0^2
+ 4 {\beta f_{\rm NL}
b_0(b_0-1) \delta_c
\over 
{\cal M}(k)D(z)}
+ {25 \over 9}
\times 0.8 \times 10^3
{\gamma (6f_{\rm NL}/5)^{4/3} \delta_c^2
(b_0 - 1)^2 \over {\cal M}(k)^2D(z)^2 }
\right]~,
\end{eqnarray}
where we have used the relation given by
Eq.~(\ref{eq:consistency}).
Hence,
it could be also possible to 
 distinguish
the "ungaussiton" model from
the other models
by making use of
LSS surveys.
\begin{figure}[htbp]
  \begin{center}
    \includegraphics{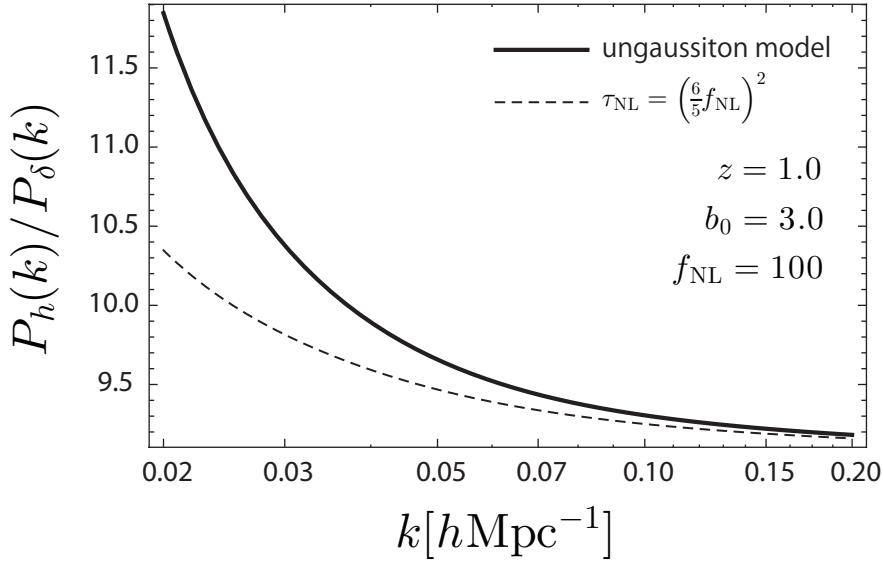}
  \end{center}
  \caption{$P_h(k)/P_\delta(k)$
  as a function of $k$ with $b_0 = 3.0$ and $f_{\rm NL} = 100$ at $z=1.0$.
  The solid line is for the "ungaussiton" model and the dashed line for the case where $\tau_{\rm NL} = (6f_{\rm NL} / 5)^2$.}
  \label{fig:fig1.eps}
\end{figure}
In Fig.~\ref{fig:fig1.eps},
we plot $P_h(k) / P_\delta(k)$ as a function of $k$
with fixing $b_0 = 3.0$, $f_{\rm NL} = 100$ and $z=1.0$.
The solid line is for the "ungaussiton" model and the dashed line for
the case where the consistency relation between $f_{\rm NL}$ and $\tau_{\rm NL}$
is given by $\tau_{\rm NL} = (6 f_{\rm NL} / 5)^2$.
From this figure, we can find 
the enhancement of $P_h(k) / P_\delta(k)$
on large scales ($ k \sim 0.01 h {\rm Mpc}^{-1}$)
in the "ungaussiton" model 
compared with the case with $\tau_{\rm NL} = (6f_{\rm NL}/5)^2$.

\section{Summary and discussion}
\label{sec:sum}

The scale-dependence of the bias of halos has been recently focused on
as a powerful tool to give a constraint on the primordial non-Gaussianity
from the LSS surveys.
In this paper, we investigate the halo bias in the ungaussiton model, which
predicts the large primordial non-Gaussianity induced from a Gaussian-squared field,
by employing the peak-background split formalism.
This model can be realized in the case where
the primordial fluctuations are sourced from both the inflaton and the curvaton
fluctuations and the curvaton stays at the origin during inflation,
and predicts the large non-Gaussianity and the specific relation
between the non-linearity parameters $f_{\rm NL}$ and $\tau_{\rm NL}$.

We calculate not only the power spectrum of the density
fluctuations of halos but also the cross power spectrum of the
matter density fluctuations and halo density fluctuations.
Then, we find that in the ungaussiton model
the effect of the non-Gaussianity on the halo bias,
which appears in the power spectrum of the halo density fluctuations, 
differs from that in the standard local-type non-Gaussianity case
due to the different consistency relation between $f_{\rm NL}$ and $\tau_{\rm NL}$.
As it is for the CMB observations,
it is expected that
the future LSS surveys  
will make us distinguish the model
from the other models where the large primordial
non-Gaussianity can be predicted. 
It is also interesting work
to check our formula by performing
the N-body numerical simulation.

\noindent {\bf Acknowledgments:}
The author would like to thank Olivier Dor{\'e} for useful comments.
This work is partially supported by the
Grant-in-Aid for Scientific research from the Ministry of Education,
Science, Sports, and Culture, Japan, No. 22340056.
The author also acknowledges support from the Grant-in-Aid for
the Global COE Program ``Quest for Fundamental Principles in the
Universe: from Particles to the Solar System and the Cosmos'' from
MEXT, Japan.

\appendix


\end{document}